\documentclass[twocolumn,showpacs,preprintnumbers,amsmath,amssymb,prl]{revtex4-1}
\usepackage{mathrsfs}
\usepackage{graphicx}
\usepackage{dcolumn}
\usepackage{bm}
\usepackage{amsmath}
\usepackage{amsfonts}
\begin{document}

\title{Topological States in Ferromagnetic CdO/EuO Quantum Well}

\author{Haijun Zhang, Jing Wang, Gang Xu, Yong Xu and Shou-Cheng Zhang}
\affiliation{ Department of Physics, McCullough Building, Stanford University, Stanford, CA 94305-4045, USA}

\begin{abstract}
Based on {\it ab-initio} calculations, we demonstrate that the ferromagnetic CdO/EuO superlattice is a simple Weyl semimetal with two linear Weyl nodes in the Brillouin zone, and the corresponding CdO/EuO quantum well realizes the stichometric quantum anomalous Hall (QAH) state without random magnetic doping. In addition, a simple effective model is presented to describe the basic mechanism of spin polarized band inversion in this system.
\end{abstract}

\date{\today}


\maketitle


The anomalous Hall effect was first discovered in ferromagnetic metals without external magnetic field~\cite{Hall1881}, which originates from the spin-orbit coupling (SOC) between the current and magnetic moments\cite{Nagaosa2010}. Recently, the quantum anomalous Hall (QAH) effect has been proposed in magnetic topological insulators\cite{qi2006,qi2008,liu2008,li2010,yu2010}, in which the anomalous Hall conductance is quantized
to $\nu e^2/h$ ($\nu$ is an integer) without the orbital magnetic field and the associated Landau levels~\cite{haldane1988}.
In magnetic topological insulators\cite{qi2006,qi2008,liu2008,li2010,yu2010}, the SOC and ferromagnetic (FM) ordering combine to give rise to the topologically nontrivial band structure. Recently, the QAH effect has been experimentally observed in Cr-doped (Bi,Sb)$_2$Te$_3$ around 30~mK~\cite{chang2013}. The chiral edge states in the QAH state conducts current without dissipation, and could be used for interconnects
of semiconductor devices\cite{zhang2012}. For potential device applications, it is important to find stichometric materials for the QAH effect, and to increase the working temperature by increasing the topologically non-trivial band gap as well as the Curie temperature ($T_c$) of magnetic moments.

Recently, the quantum spin Hall (QSH) effect was theoretically proposed\cite{Liu2008a} and experimentally observed in the broken-gap type-II InAs/GaSb quantum well (QW)\cite{knez2011,du2013}. In this system, topological insulator property arises from the relative band inversion between the two different materials of the quantum well\cite{Liu2008a}. In this Letter, we show that the basic mechanism of relative band inversion can also
be applied to oxide heterojunctions, opening a new material class for the study of topological states. As an illustrative example,
we propose to realize the QAH state in a broken-gap type-II (CdO)$_n$(EuO)$_m$ QW, in which a large band gap and high $T_c$ of FM order can be achieved. Here, ``$n$'' and ``$m$'' refer to the number of atomic layers of (CdO) and (EuO),respectively. As we know, EuO is an important FM insulator where the ferromagnetism is contributed by transition metal element Eu and its $T_c$ is around 69~K. In addition, EuO thin films has similar $T_c$ as the bulk one~\cite{muller2009}. A high $T_c$ is thus expected for (CdO)$_n$(EuO)$_m$ QW as well. Also, the band gap of the QW system can be tuned by changing the layer thickness $(n,m)$. All these properties in the (CdO)$_n$(EuO)$_m$ QW are distinct from the previous proposals in magnetically
doped topological insulators, such as Mn-doped HgTe~\cite{liu2008} and Cr-doped Bi$_2$Se$_3$-family~\cite{yu2010}, where a high $T_c$ and a large band gap are hard to achieve simultaneously. Moreover, the high quality epitaxial growth of EuO~\cite{Iwata2000,Miyazaki2009,Sutarto2009,Swartz2010,Ulbricht2008,Swartz2012} and CdO~\cite{yan2001,jefferson2008} have been reported by many groups using various techniques, including pulsed laser deposition (PLD) and metal-organic vapor-phase epitaxy (MOVPE). The (CdO)$_n$(EuO)$_m$ QW is a promising system to realize the QAH effect with both a high $T_c$ and a tunable band gap.

\begin{figure}[t]
\begin{center}
\includegraphics[width=2.5in,clip=true,angle=-90]{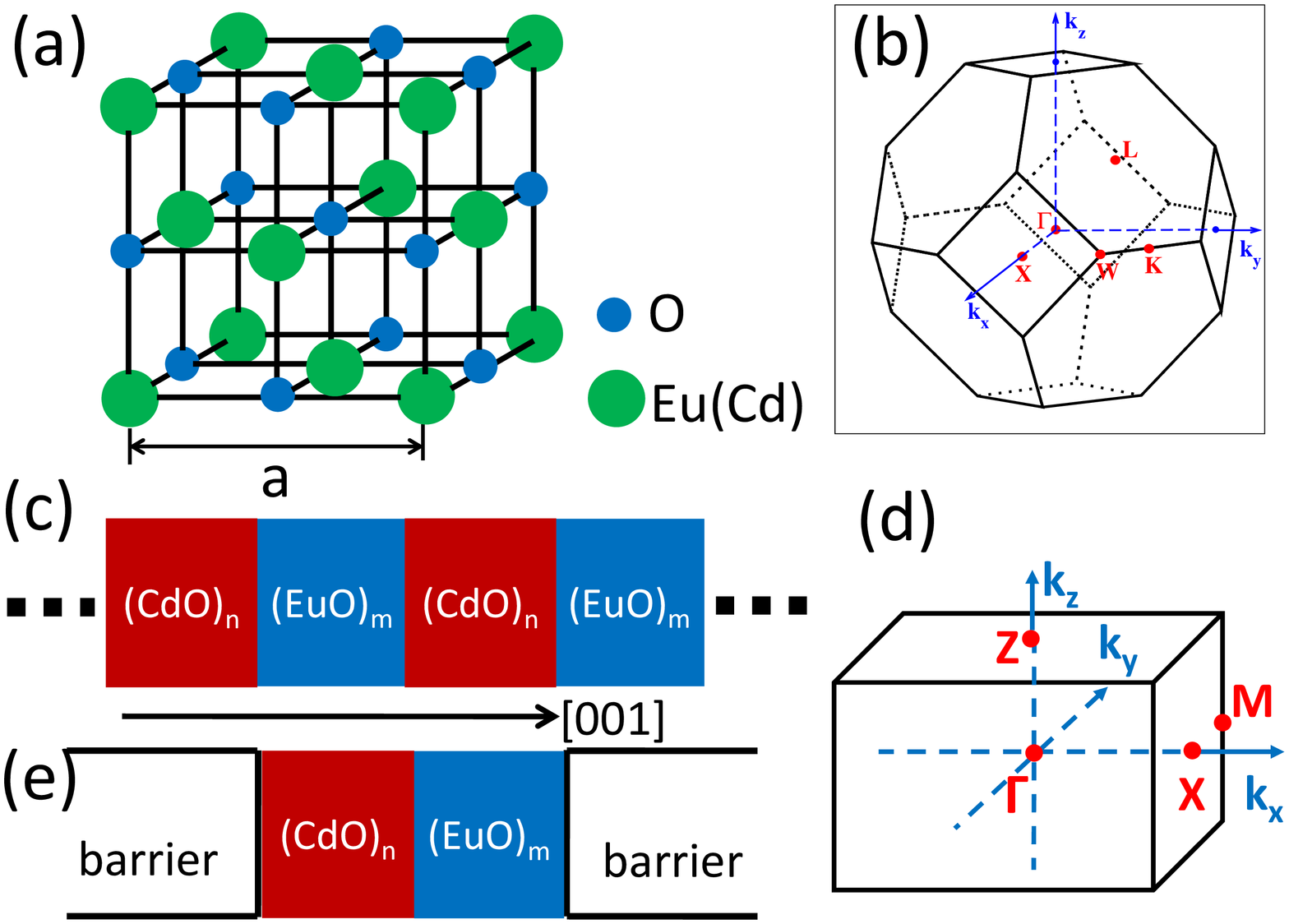}
\end{center}
\caption{(color online). (a) The rock-salt crystal structure of CdO and EuO with $F{m\bar{3}m}$ space group (No. 225). (b) The Brillouin zone (BZ) for the rock-salt structure. The TRIM points are marked, for example, $\Gamma(0,0,0)$, $X(\pi,\pi,0)$ and $L(\pi,0,0)$.(c) The schematic of (CdO)$_n$(EuO)$_m$ superlattice. (d) The BZ of (CdO)$_n$(EuO)$_m$ superlattice. (e) The schematic of (CdO)$_n$(EuO)$_m$ quantum Well.}\label{fig1}
\end{figure}

\begin{figure*}[t]
   \begin{center}
      \includegraphics[width=4.5in,angle=-90]{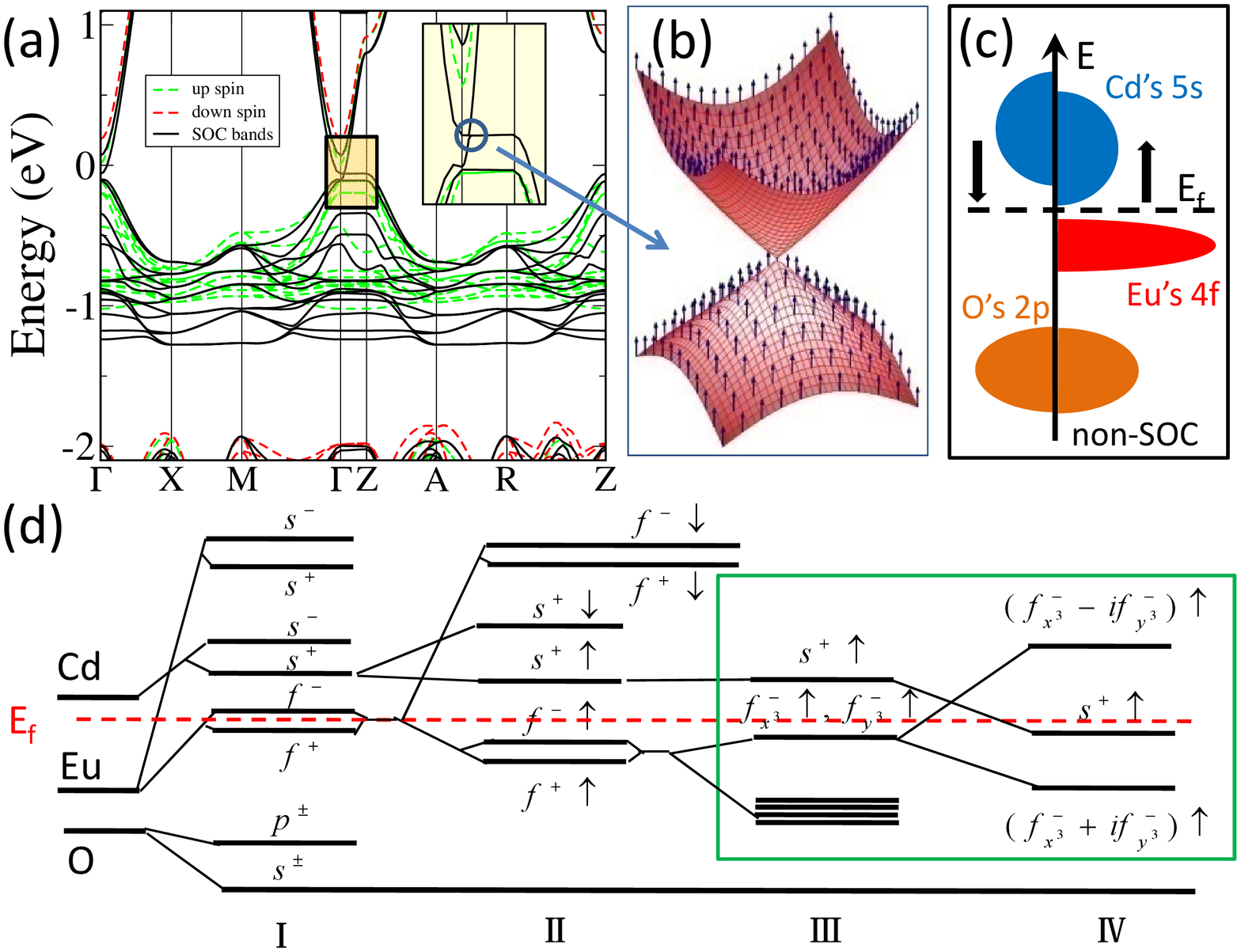}
    \end{center}
    \caption{(color online).
    (a) Band structure of the (CdO)$_2$(EuO)$_2$ superlattice with and without SOC. The green/red dashed lines denote the up-spin/down-spin bands. The black solid lines represent the SOC bands. There is a band inversion with the SOC bands. The bands around Fermi level at $\Gamma$ is zoomed in the inset. (b) The Weyl nodes with the spin texture at ($0,0,k_c$). (c) The schematic of density of states (DOS) without SOC. The right/left part is the up-spin/down-spin DOS.(d) Evolution of the band sequence at $\Gamma$ starting from the atomic levels. Four stages (I-IV) represent turning on the chemical bonding (I), the exchange field (II), the crystal field (III) and the SOC effect, in sequence. The red dashed is the Fermi level.
     }
    \label{fig2}
\end{figure*}

Both bulk CdO and EuO have a rock-salt structure with $F{m\bar{3}m}$ space group (No.~225), where each of the two atoms in the unit cell forms a separate face-centered cubic lattice with the two lattices interpenetrating, as shown in Fig.~1(a). The lattice constants of CdO and EuO are $4.69$~\AA \cite{yan2001} and $5.14$~\AA,\cite{Sutarto2009} respectively. The lattice mismatch between EuO and CdO can be reduced by Ba doping in CdO (for example, $4.98$\AA for Ba$_{0.15}$Cd$_{0.85}$O)~\cite{Vazhnov1989}. Because the Ba doping in CdO essentially doesn't change the electronic structure,\cite{supp} for simplicity, we neglect the Ba doping effect and focus on pure CdO and EuO. The Brillouin zone (BZ) of the rock-salt structure has eight time-reversal-invariant-momentum (TRIM) points, 1$\Gamma(0,0,0)$,3$X(\pi,\pi,0)$ and 4$L(\pi,0,0)$ [see Fig.~1(b)]. By alternately stacking (CdO)$_n$ and (EuO)$_m$ along [001] direction, we can construct a superlattice [see Fig.~1(c)]. The BZ becomes the simple-tetragonal lattice, as shown in Fig.~1(d). The (CdO)$_n$(EuO)$_m$ QW structure can be constructed by taking one basic unit [(CdO)$_n$(EuO)$_m$] of the superlattice into insulating barriers (e.g. SrO) [see Fig.~1(e)].

The Vienna Ab-initio Simulation Package (VASP)\cite{kresse1993,kresse1999} is employed to carry out {\it ab-initio} calculations with the framework of the Perdew-Burke-Ernzerhof-type\cite{perdew1996} generalized gradient approximation of density functional theory\cite{hohenberg1964}. Projector augmented wave\cite{blochl1994} pseudo-potentials are used. 16$\times$16$\times$16 and 10$\times$10$\times$6/10$\times$10$\times$2 are used for the k-mesh of bulk and superlattice/QW calculations, respectively. The kinetic energy cutoff is fixed to 450eV. The lattice constant and internal atomic positions are fully relaxed in superlattices with the force cutoff 0.01$eV/$\AA. In order to study the electronic structure of QW structure, we take SrO as the barrier material which has the same lattice constant with that of EuO, so we fixed the lattice constant of EuO and relaxed all the atoms. In QWs, [(CdO)$_n$(EuO)$_m$] and the nearest SrO layers are relaxed. We add the effective Hubbard U$_s$(14~eV) on Cd's $s$ orbital and U$_f$ (7.5~eV) on Eu's $f$ orbitals by employing GGA+U method to fit experiments, which is crucial to obtain the correct band alignment between CdO and EuO in the superlattice and QW structures. Spin-orbit coupling (SOC) effect is included with a non self-consistent calculation.


The insulating state of CdO with a direct band gap (2.2~eV) is predicted by GGA+U calculations, which matches the optical measurement~\cite{jefferson2008}, and the valence bands hold the same feature as those from GGA calculations. GGA+U calculations also predicts an insulating state for EuO with a direct band gap (around 0.8~eV) at X points which is slightly smaller than the experimental value (1.1~eV).\cite{Mauger1986}.

The (CdO)$_n$(EuO)$_m$ superlattices have qualitatively the same electronic structure. We take the (CdO)$_2$(EuO)$_2$ superlattice as an example for the following discussion. The FM state is the ground state with a large local moment (around 7$\mu_B$) on each Eu atom according to Hunds' rule. Based on ab-initio calculations, we find that a compressive strain enables the [001] direction as the easy magnetization axis, so the magnetic moment is  fixed along the [001] direction in this work. The band structures of the (CdO)$_2$(EuO)$_2$ superlattice with and without SOC are shown in Fig.~2(a). The bottom of the conduction bands is contributed by Cd's $s$ orbital, and the top of the valence bands is Eu's $|f,\uparrow\rangle$ orbital. Due to the $s$-$f$ exchange interaction, the conduction bands are spin split giving rise to two spin polarized bands $|s,\uparrow\rangle$ and $|s,\downarrow\rangle$. After SOC is included, the $|s,\uparrow\rangle$ band crosses $|f,\uparrow\rangle$ and gives rise to a band inversion at $\Gamma$ which is topologically non-trivial due to the different symmetries of $|s,\uparrow\rangle$ (even parity) and $|f,\uparrow\rangle$ (odd parity) [see the inset of Fig.~2(a)]. The schematic density of states (DOS) without SOC is shown in Fig.~2(c). One Weyl node, discussed in the following, is seen at $(0,0,k_c)$ between $\Gamma(0,0,0)$ and $Z(0,0,\pi)$ [see Fig.~2(b)]. It has a linear dispersion, and the out-of-plane spin points to `$+z$' direction, as shown in Fig~2(b).

\begin{figure}[t]
   \begin{center}
      \includegraphics[width=2.5in,angle=-90]{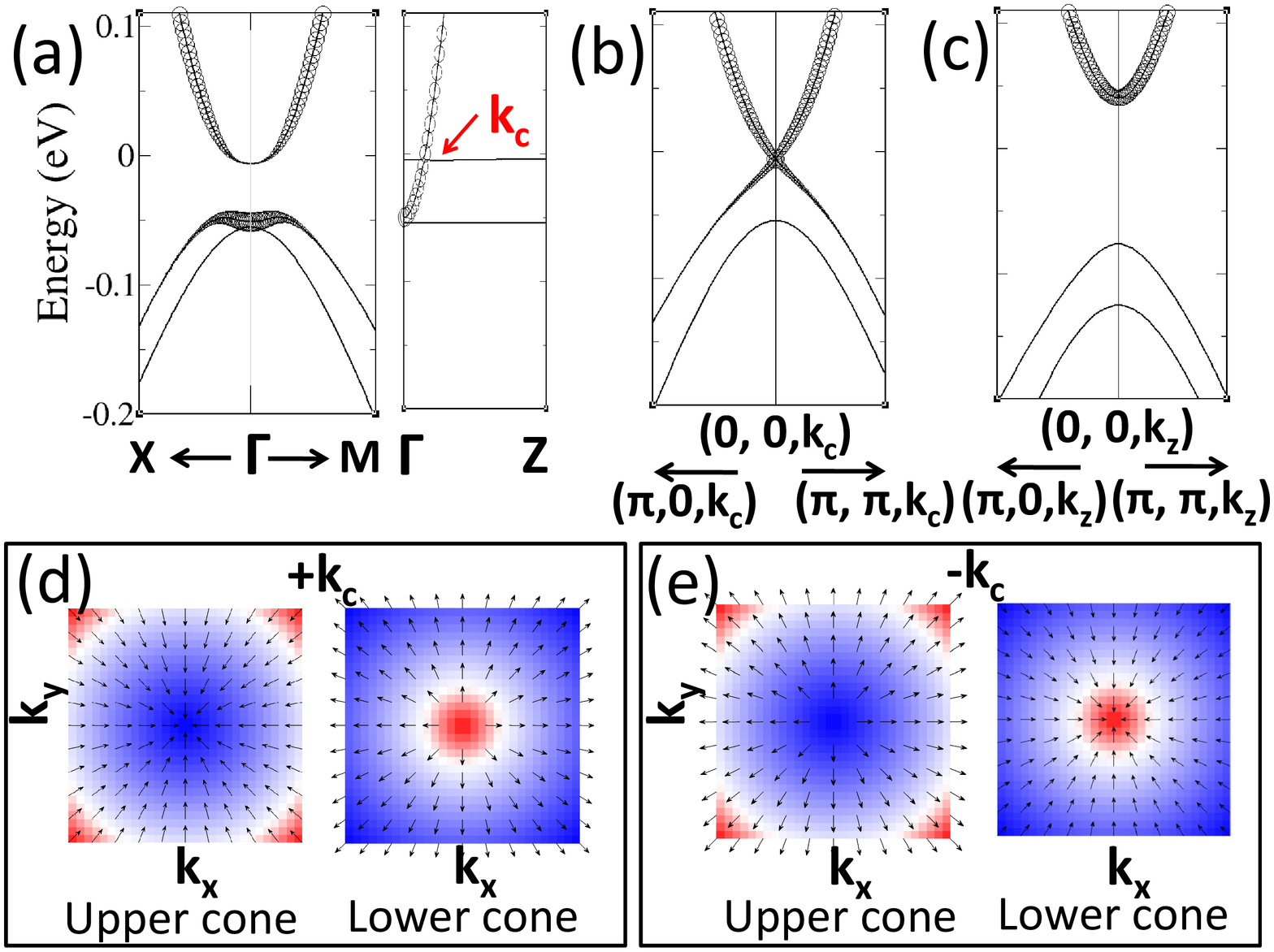}
    \end{center}
    \caption{(color online).
    (a-c) The fat band of the (CdO)$_2$(EuO)$_2$ superlattice for the $\Gamma$, $(0,0,k_c)$ and $(0,0,k_z)$ planes. The size of black circles means Cd's $s$-orbital projection. A band inversion can be observed at $\Gamma$ in (a). A Weyl node locates at $(0,0,k_c)$ (b). $(0,0,k_z)$ is one point between $(0,0,k_c)$ and $Z(0,0,\pi)$. (d-e) The in-plane spin texture of two Weyl nodes at $(0,0,\pm k_c)$. The left figure of (d) and (e) is for upper Weyl cone, and the right one of (d) and (e) is for lower Weyl cone. The flow direction is exactly opposite for these two Weyls nodes.
     }
    \label{fig3}
\end{figure}

To get a better understanding of the band inversion, we study the evolution of the band sequence at $\Gamma$ point by starting from the atomic energy levels [see in Fig.~2(d)]. The atomic energy levels of Cd ($5s^2$), Eu ($4f^76s^2$) and O ($2s^22p^4$) are taken as the starting point. At stage I, the chemical bonding is considered. All the atomic levels ($s$, $p$ and $f$) split into the bonding ($|s^+\rangle$, $|p^+\rangle$ and $|f^+\rangle$) and antibonding ($|s^-\rangle$, $|p^-\rangle$ and $|f^-\rangle$) levels, where `$+(-)$' represents the even (odd) parity. The energy levels $|s^{\pm}\rangle$ and $|p^{\pm}\rangle$ of O atoms stay deep below the Fermi level, and $|s^{\pm}\rangle$ of Cd and Eu atoms are above Fermi level. The energy levels $|f^{\pm}\rangle$ of Eu atoms cross Fermi level without bonding with other levels. At stage II, the exchange interaction is considered. The energy levels $|f^{\pm}\rangle$ of Eu atoms split into the low energy levels $|f^{\pm}\uparrow\rangle$ and the high energy levels $|f^{\pm}\downarrow\rangle$ due to Hund's rule. With the $s$-$f$ exchange interaction, the energy levels $|s^{\pm}\rangle$ split into the high energy levels $|s^{\pm}\downarrow\rangle$ and the low energy levels $|s^{\pm}\uparrow\rangle$. At stage III, the crystal field is turned on. The energy levels $|f^{\pm}\uparrow(\downarrow)\rangle$ split into different groups. Especially, the energy levels $|f^{-}_{x^3}\uparrow\rangle$ and $|f^{-}_{y^3}\uparrow\rangle$ are degenerate just below Fermi level, while $|s^+\uparrow\rangle$ is just above Fermi level. At stage IV, SOC effect is considered. The degeneracy of $|f^{-}_{x^3}\uparrow\rangle$ and $|f^{-}_{y^3}\uparrow\rangle$ is broken and they recombine into the high energy level ($|f_-^-\uparrow\rangle$=$|f^{-}_{x^3}\uparrow\rangle-i|f^{-}_{y^3}\uparrow\rangle$) and the low energy level ($|f_+^-\uparrow\rangle$=$|f^{-}_{x^3}\uparrow\rangle+i|f^{-}_{y^3}\uparrow\rangle$). With strong enough SOC, a band inversion occurs between $|f_-^-\uparrow\rangle$ and $|s^+\uparrow\rangle$, which is topologically non-trivial due to opposite parities.

If taking $k_z\in \{-\pi\sim\pi\}$ as a parameter, we can evaluate Chern number $\cal C$ for the band structure of the $k_z$ plane. ${\cal C}=1$ is for the plane of $|k_z|<k_c$ due to the band inversion of $|f_-^-\uparrow\rangle$ and $|s^+\uparrow\rangle$, and ${\cal C}=0$ is for the plane of $|k_z|>k_c$, as shown in Fig.~3(a) and (c). Therefore topologically unavoidable Weyl node has to exist at the phase boundary between ${\cal C}=1$ and ${\cal C}=0$ planes (i.e, at $k_z=\pm k_c$).\cite{Xu2011a} Weyl fermion is described by $2\times2$ Pauli matrices, which is considered as half of Dirac fermion, so there is no way to open an energy gap on Weyl fermion.\cite{Wan2011} In our case, there are two Weyl nodes locating at $(0,0,\pm k_c)$ with the in-plane spin textures due to the SOC effect, as shown in Fig.~3(d) and (e). At $(0,0,+k_c)$, all spins flow into the center for the upper Weyl cone, and all spins flow out of the center for lower Weyl cone. The situation is exactly opposite to the Weyl node at $(0,0,-k_c)$, which presents the opposite chirality.


Based on the analysis of the band sequence discussed above, a simple $2\times2$ model, with $|f_-^-,\uparrow\rangle$ and $|s^+,\uparrow\rangle$ as the basis, can be introduced to describe the spin polarized band inversion:
\begin{equation}
\mathcal{H}_{\mathrm{eff}} =\begin{pmatrix}
M & Dk_-\\
Dk_+ & -M
\end{pmatrix},
\end{equation}
where $k_{\pm}=k_x\pm i k_y$, and $M=M_0-\beta_1(k_x^2+k_y^2)-\beta_2k_z^2$ is the mass term expanded to the second order, with parameters $M_0>0$ and $\beta_1, \beta_2>0$ to ensure the band inversion. Since the two bases have opposite parity, the off-diagonal element has to be odd in $k$. In addition, the $k_{\pm}$ has to appear to conserve the angular moment along the $z$ direction. Therefore, to the leading order, the $k_\pm$ is the only possible form for the off-diagonal element. Evaluating the eigenvalues $E(k)=\pm\sqrt{M^2+D^2(k_x^2+k_y^2)}$
, we get only two gapless solutions at $k_z=\pm k_c=\pm\sqrt{M_0/\beta_2}$ which are exactly the two Weyl nodes obtained from our {\it ab-initio} calculations, as shown in the inset of Fig.~2(a). Because of the presence of $k_\pm$ in the off-diagonal elements, it is easy to check that Chern number ${\cal C}$ equals 1 for the planes with $-k_c<k_z<k_c$. The band dispersion near the Weyl nodes [see Fig.~3(b)] is thus linear, with a chiral in-plane spin texture [see Fig.~3(d)]. The two Weyl nodes have opposite chirality due to the opposite sign of the mass term, and they form a single pair of magnetic monopoles carrying the gauge flux in $\vec{k}$ space.

\begin{figure}[t]
   \begin{center}
     \includegraphics[width=2.8in,angle=-90]{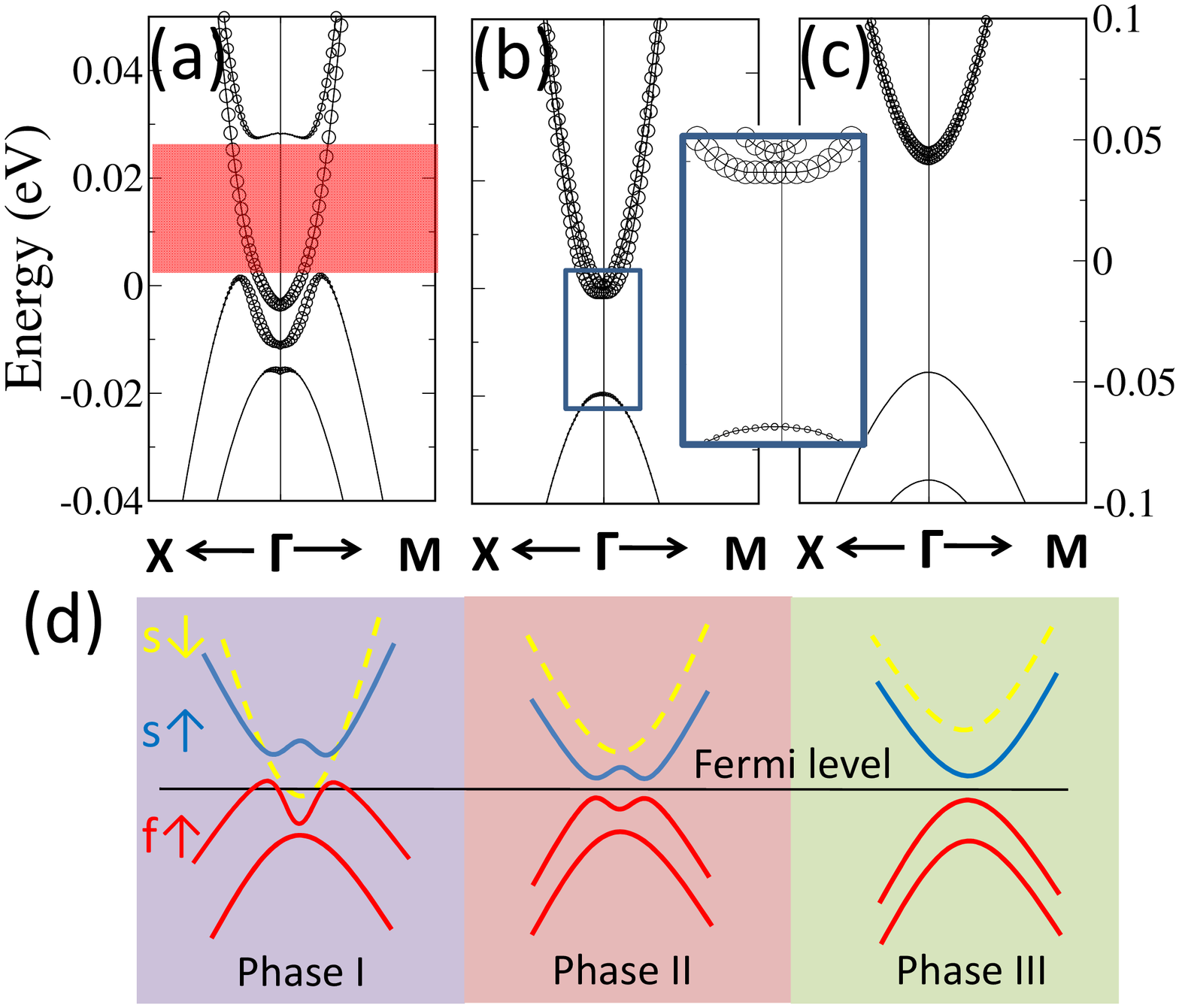}
    \end{center}
    \caption{(color online). (a) Band structure of the (CdO)$_4$(EuO)$_3$ quantum Well. The size of black circles means Cd's $s$-orbital projection. The two conduction bands $|s^{+}\uparrow\downarrow\rangle$ both cross the valence band $|f_-^-\uparrow\rangle$. The red region presents a half-metal range with a single Fermi surface. (b) Band structure of the (CdO)$_4$(EuO)$_3$ quantum Well with the enlarged in-plane lattice constant by 1 \%. A full band gap is opened with a band inversion between the $|s^{+}\uparrow\rangle$ and $|f_-^-\uparrow\rangle$ bands, which presents a QAH phase. The band structure is zoomed in in the inset. (c) Band structure of  the (CdO)$_4$(EuO)$_3$ quantum Well with the enlarged in-plane lattice constant by 2 \%. No band inversion happens between the bands $|s^{+}\uparrow\downarrow\rangle$ and $|f_-^-\uparrow\rangle$, which presents a trivial insulator phase. (d) The schematic of these three different phases in (a-c).
     }
    \label{fig4}
\end{figure}

Whereas the Weyl semi-metal state can be realized in the superlattice structure, the QAH effect can be realized in the (CdO)$_n$(EuO)$_m$ QW structure with appropriate barrier layers. The band structure of the QW is qualitatively the same to that of the $k_z=0$ plane in the corresponding superlattice. Here we show the numerical results on the (CdO)$_4$(EuO)$_3$ QW. Both Cd's $|s^+\uparrow\downarrow\rangle$ bands cross Eu's $|f_-^-\uparrow\rangle$ band to present a half-metal phase with a single Fermi surface, named phase I, and its band structure is shown in Fig.~4(a) where the single-Fermi-surface range is marked in red. This half-metal phase is no other than the candidates for Majorana fermion by the proximity with s-wave superconductors~\cite{Chung2011}. With the increased in-plane lattice constant by 1\%, Cd's $|s^+\downarrow\rangle$ band becomes unoccupied and stay above Eu's $|f_-^-\uparrow\rangle$ band. In this case, the QAH effect, in which the bulk band gap is around 20meV, is expected to be realized due to the band inversion between Cd's $|s^+\uparrow\rangle$ and Eu's $|f_-^-\uparrow\rangle$ bands, named phase II, shown in Fig.~4(b). If the in-plane lattice constant is enlarged more by 2\%, both Cd's $|s^+\uparrow\downarrow\rangle$ bands have no band inversion with Eu's $|f_-^-\uparrow\rangle$ band, and the system becomes a trivial insulator, named phase III, shown in Fig.~4(c). Fig.~4(d) shows the schematic of all the phases. The enlarged in-plane lattice constant can be realized by using BaO or Ba-doped SrO as the substrate in experiments.

 We stress that the topological effects discussed in this letter can occur within a large parameter space. Even if the discussed effects may not occur with the parameters exactly the same as our {\it ab-initio} calculations, they can still be recovered by changing a few other parameters in experiments. For example, in addition to the in-plane lattice constant, the thickness of the (CdO)$_n$ layer is another degree of freedom to tune the band inversion between Cd's $s$ band and Eu's $f$ band. The (CdO)$_3$(EuO)$_2$ QW is shown as another example in our online supplementary materials.\cite{supp} When the thickness of the (CdO)$_n$ layer is reduced, the size effect tends to reduce this band inversion. Oppositely the band inversion will be enhanced by increasing the thickness. Moreover, a p-n junction could be formulated with n-doped (CdO)$_n$ and p-doped (EuO)$_m$, in which a built-in voltage can be induced to assist the band inversion. Also the external electric gating on (CdO)$_n$ layer can increase or reduce that built-in voltage. In summary, the CdO/EuO structure is a highly adjustable platform which is promising to search for topological states, including Weyl semimetal and the QAH state, which are more favorable for applications.

\begin{acknowledgments}
This work is supported by US Department of Energy, Office of Basic Energy Sciences, Division of Materials
Sciences and Engineering, under contract DE-AC02-76SF00515 and by FAME, one of six centers of STARnet, a
Semiconductor Research Corporation program sponsored by MARCO and DARPA.

\end{acknowledgments}
%



%

\end{document}